\documentclass{elsart}

\usepackage{graphics}
\usepackage{graphicx}

\usepackage{amssymb}

\begin{document}

\begin{frontmatter}



\title{Quarkonium Production in PHENIX}


\author{Abigail Bickley, for the PHENIX Collaboration \cite{Collab}} 
\address{
\centerline{Department of Physics, 390 UCB}
\centerline{University of Colorado, Boulder, Colorado, 80309, USA}
}

\begin{abstract}
Quarkonia provide a sensitive probe of the properties of the hot dense medium created in high energy heavy ion collisions.  Hard scattering processes result in the production of heavy quark pairs that interact with the collision medium during hadronization.  These in medium interactions convey information about the fundamental properties of the medium itself and can be used to examine the modification of the QCD confining potential in the collision environment.  Baseline measurements from the d+Au and p+p collision systems can be used to distinguish cold nuclear matter effects while measurements from heavy ion collision systems, Au+Au and Cu+Cu, can be used to quantify in-medium effects.  PHENIX results for the production of the $J/\psi$ for a diverse set of collision systems and energies and for the $\Upsilon$ in p+p collisions are presented.  
\end{abstract}

\begin{keyword}

\PACS 
\end{keyword}
\end{frontmatter}

\section{Introduction}
Quarkonia are predominately produced in hadronic collisions through hard scattering processes that result in the formation of heavy quark pairs.  These pairs interact with the collision medium and convey information about its fundamental properties.  However, a number of competing effects such as color screening, recombination, gluon dissociation, shadowing and normal nuclear absorption influence the observed quarkonia yields.  Experimental comparisons of these yields in a wide range of collisions systems can be used to resolve the relative importance of each effect.  In heavy ion collisions quarkonia are used as probes of the conditions achieved in the hot and dense medium,  whereas initial state and nuclear medium effects can be disentangled through studies of p+p and d+Au collisions.  The PHENIX experiment at the Relativistic Heavy Ion Collider has studied the production of the $J/\psi$ in p+p, d+Au, Cu+Cu and Au+Au collisions and has shown preliminary analyses of the production of the $\Upsilon$.  This extensive set of data allows a direct comparison to be made between cold and hot nuclear systems.  

\section{Theory Overview}
The suppression of $J/\psi$ production in the collision medium was originally proposed in \cite{MatsuiSatz} to be a signature of deconfinement.  This prediction was based upon the premise that heavy quarkonia are only formed during the initial hard nucleon-nucleon collisions and any subsequent interactions result in the additional loss of yield.  Suppression was expected to occur in deconfined systems through the mechanism of color screening which prevents $c\bar{c}$ binding in the interaction region if the screening radius is smaller than the $J/\psi$ radius.  However, the measured $J/\psi$ yield contains contributions from both direct production and feed-down from higher resonance states such as the $\chi_{c}$ and $\psi'$.  The relative particle yield from each source has been experimentally found to be approximately $60\%$ direct production, $30\%$ feed-down from the $\chi_{c}$ state and $10\%$ $\psi'$ feed-down \cite{HERA-B}.  Each of these resonance states can only exist in the collision medium as long as the color screening length is larger than the radius of the state.  Thus the environmental conditions present in the collision determine whether the states exists in the bound form and contribute to the measured $J/\psi$ yield.  The sequential melting of the $c\bar{c}$ states could serve as a thermodynamic probe of the medium \cite{satz_seq} provided that their melting points are sufficiently different to be resolved experimentally.  While the melting points of the $\psi'$ and $\chi_{c}$ are expected to occur close to the critical temperature, recent lattice calculations indicate that the $J/\psi$ could exist in the bound state up to temperatures of approximately $2T_{c}$ \cite{lattice}.  Alternatively, recombination models provide a competing mechanism to color screening since they include the formation of $J/\psi$ from independently produced c and $\bar{c}$ quarks.  Experimentally this could lead to an enhancement (or less dramatic suppression) of the measured $J/\psi$ yield. 

\section{PHENIX Experiment}
The PHENIX experiment is designed to study quarkonia via their di-lepton decay channels \cite{Collab}.  At mid-rapidity, $|\eta|<0.35$, the ring imaging \v{C}erenkov detectors, drift chambers and electromagnetic calorimeters are used to detect $J/\psi \rightarrow e^{+}e^{-}$ decays.  The muon detectors, consisting of cathode strip tracking chambers in a magnetic field and Iarocci tube planes with thick steel absorbers, are used to measure $J/\psi \rightarrow \mu^{+}\mu^{-}$ and $\Upsilon \rightarrow \mu^{+}\mu^{-}$ at forward and backward rapidities, $1.2 < |\eta| < 2.2$.  The quarkonia measurements shown result from sampling $0.35~pb^{-1}$ p+p data, $2.74~nb^{-1}$ d+Au data, $241~\mu b^{-1}$ Au+Au data and $3.0~nb^{-1}$ Cu+Cu data.  The quarkonia yields are extracted from the raw data by determining the number of unlike-sign di-lepton pairs that are found within the $J/\psi$ (or $\Upsilon$) mass window after background subtraction via event mixing or like-sign subtraction.  All yields are corrected for the acceptance and efficiency of the detector elements and the triggering efficiency, and then normalized by the number of recorded collisions.  

\section{$J/\psi$ Production in p+p and d+Au Collisions}
\begin{figure}
\begin{center}
\includegraphics[width=4.0in]{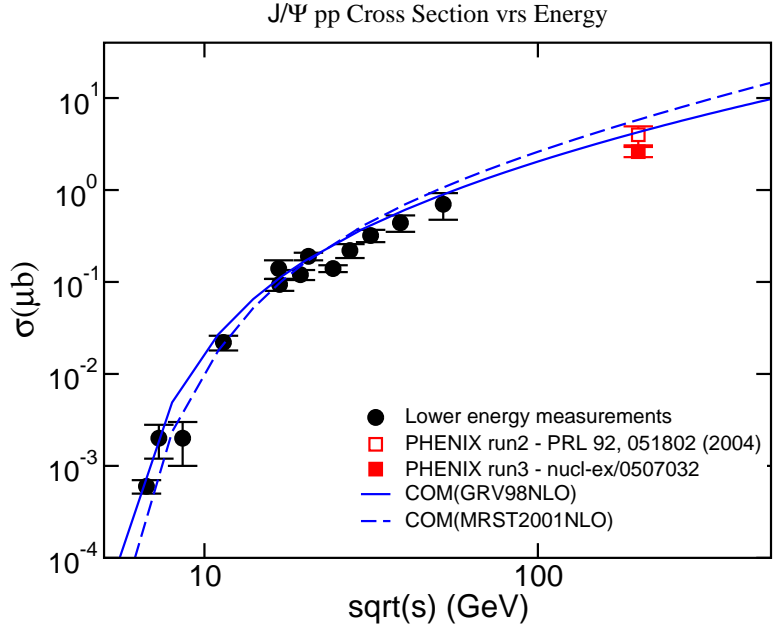}
\end{center}
\caption[jpsi_roots]{$J/\psi$ cross section in p+p collisions as a function of collision energy.\cite{ppg17}} 
\label{jpsi_roots}
\end{figure}

\begin{figure}
\begin{center}
\includegraphics[width=4.0in]{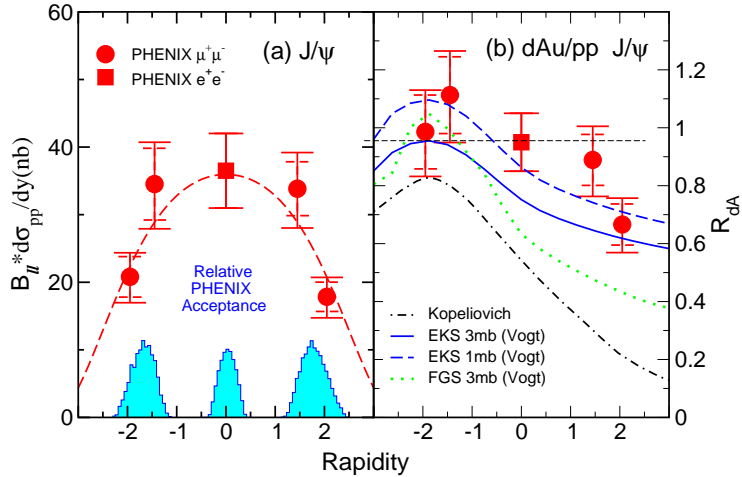}
\end{center}
\caption[sigpp_rda]{a) J/$\psi$ differential cross section times di-lepton branching ratio in p+p collisions versus rapidity; b) $J/\psi$ $R_{dA}$ versus rapidity in d+Au collisions. \cite{ppg38}} 
\label{sigpp_rda}
\end{figure}

The measurement of $J/\psi$ production in p+p collisions provides an important baseline to which all other collision systems can be compared.  The total $J/\psi$ cross section measured by PHENIX in $200~GeV$ collisions \cite{ppg17} is the highest energy measurement made to date and is consistent with the trend observed in lower energy data and predicted by the Color Octet Model, Fig.~\ref{jpsi_roots}.  The dependence of the cross section on collision energy is sensitive to the parton distribution function.  However, the existing data do not allow a distinction to be made between the available parameterizations.  Figure~\ref{sigpp_rda} shows the rapidity dependence of the differential cross section times di-lepton branching ratio.  Good agreement is found with the PYTHIA shape calculation.  

In d+Au collisions the $J/\psi$ yield can be used to distinguish the relative influence of initial state, final state and nuclear effects.  The wide rapidity coverage of the PHENIX experiment allows the modification of the parton distribution function to be probed in both the anti-shadowing and shadowing regimes.   The forward muon spectrometer, $1.2< \eta < 2.2$, covers the shadowing region, the backward muon spectrometer, $-2.2<\eta< -1.2$, lies within the anti-shadowing domain, and the central tracking detectors, $|\eta|< 0.35$,  bridge these two regimes.  

To examine the $J/\psi$ yield produced in collisions of nucleus A with nucleus B relative to p+p collisions the nuclear modification factor, $R_{AB}$, is defined as:

\begin{equation}
R_{AB} = \frac{dN_{AB}^{J/\psi}}
			{\langle Ncoll \rangle*dN_{pp}^{J/\psi}}
\end{equation}

Figure~\ref{sigpp_rda} shows the $J/\psi$ nuclear modification factor in $200~GeV$ d+Au collisions versus rapidity measured in PHENIX compared to theoretical predictions \cite{ppg38}.  The data are consistent with weak gluon shadowing and a $c\bar{c}$ absorption cross section of $1~mb$, although this value is not strongly constrained.

\section{$J/\psi$ Production in Heavy Ions Collisions}
\begin{figure}
\begin{center}
\includegraphics[width=4.0in]{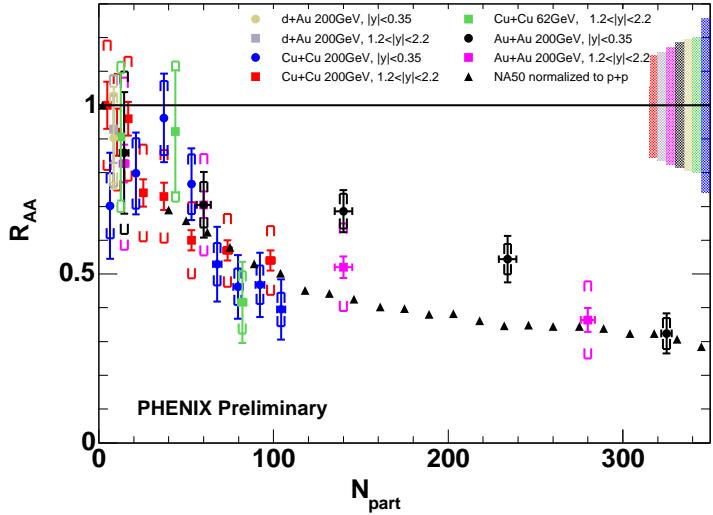}
\end{center}
\caption[hi_jpsi_cent]{$J/\psi$ $R_{AA}$ versus number of collision participants for d+Au, Cu+Cu and Au+Au collisions at mid and forward rapidities at $\sqrt{s} = 200~GeV$ and Cu+Cu collisions at forward rapidity at $\sqrt{s} = 62~GeV$. \cite{hugo}} 
\label{hi_jpsi_cent}
\end{figure}

In heavy ion collisions $J/\psi$ production acts as a probe of the nuclear effects.  Comparisons of the nuclear modification factor as a function of collision centrality, energy, rapidity and system size are possible with preliminary PHENIX Au+Au and Cu+Cu data.  Figure~\ref{hi_jpsi_cent} shows the nuclear modification factor for Au+Au, Cu+Cu and d+Au data as a function of centrality at mid and forward rapidities.  In the most central Au+Au collisions a factor of 3 suppression is observed while in Cu+Cu collisions the suppression is approximately a factor of 2.  Comparing Au+Au and Cu+Cu collisions with the same number of participants, $N_{part}$, the observed suppression is similar between the two systems.  In addition, within both Au+Au and Cu+Cu collisions the nuclear modification is similar at mid and forward rapidities.  Finally, a comparison of $62~GeV$ and $200~GeV$ Cu+Cu collisions at forward rapidity remarkably shows the same degree of suppression within the large experimental uncertainties despite the difference in collision energy.  

\begin{figure}
\begin{center}
\includegraphics[width=5.0in]{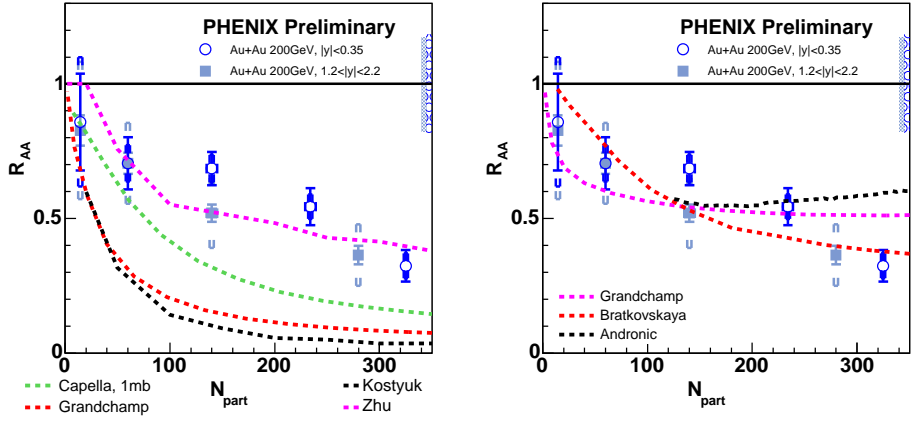}
\end{center}
\caption[hi_jpsi_y]{$J/\psi$ $R_{AA}$ versus number of collision participants for Au+Au collisions at mid and forward rapidities at $\sqrt{s} = 200~GeV$ compared to predictions from final state suppression models without (a) and with (b) recombination. \cite{hugo}} 
\label{raa_theory}
\end{figure}

\begin{figure}
\begin{center}
$\begin{array}{c@{\hspace{0.25in}}c}
\includegraphics[width=2.4in]{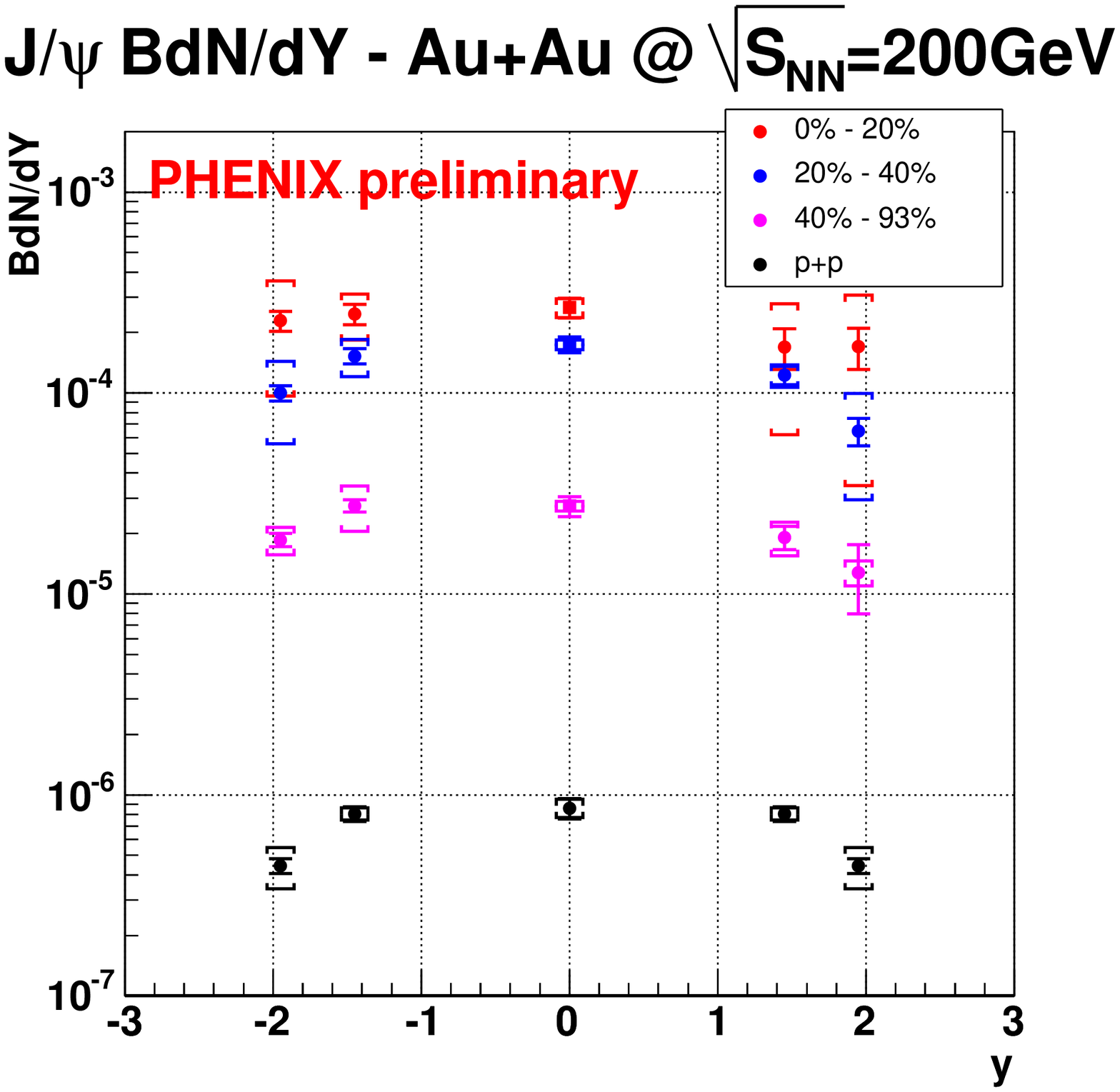}&
\includegraphics[width=2.4in]{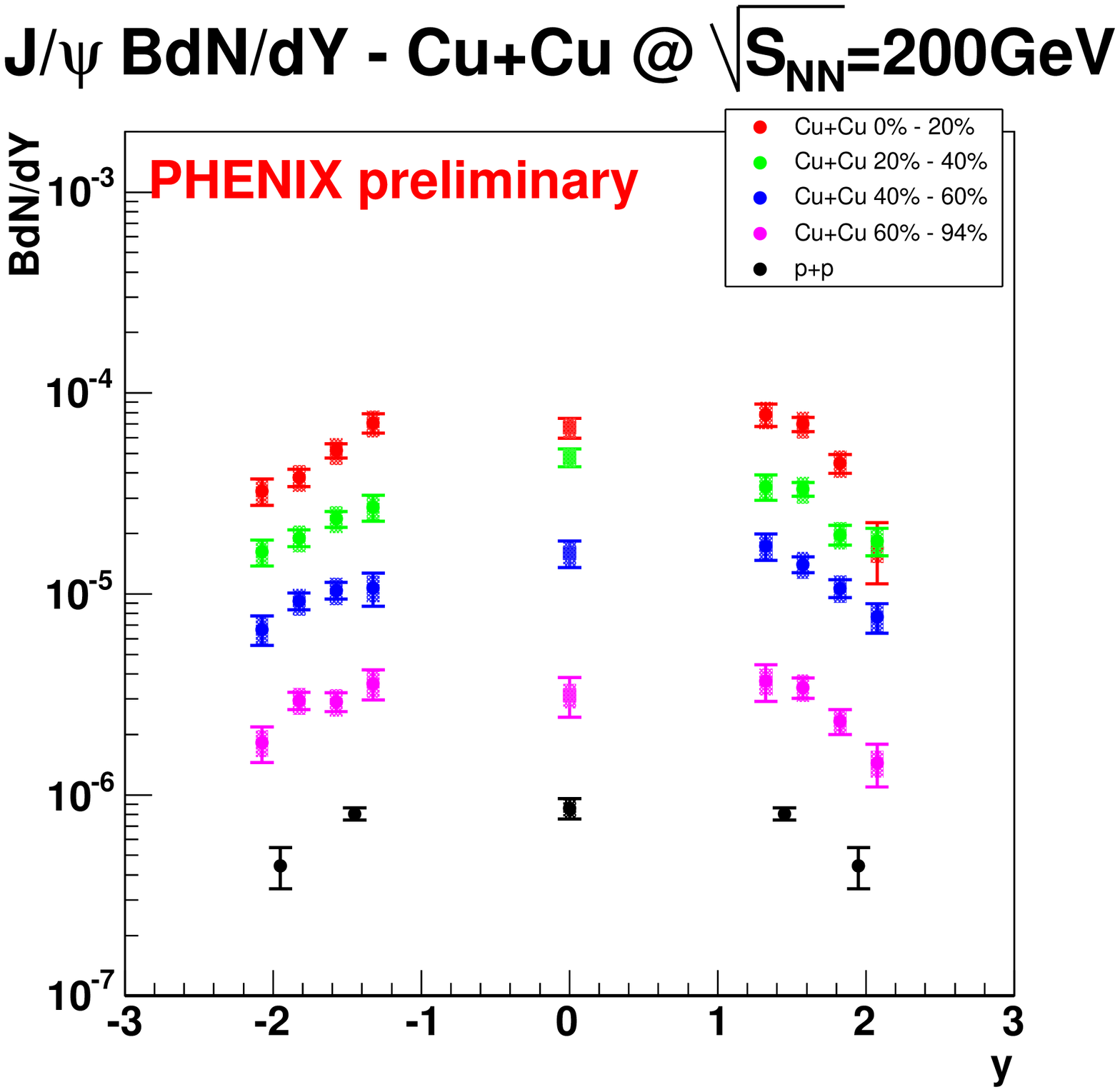}
\end{array}$
\end{center}
\caption[hi_jpsi_y]{$J/\psi$ invariant yield as a function of rapidity and centrality in $\sqrt{s} = 200~GeV$ Au+Au and Cu+Cu collisions. \cite{hugo}} 
\label{hi_jpsi_y}
\end{figure}

It is instructive to compare the PHENIX results with model calculations that include the competing effects that are thought to govern $J/\psi$ production.  Cold nuclear matter calculations \cite{Vogt} that assume a $c\bar{c}$ absorption cross section of $1~mb$, which is consistent with the PHENIX d+Au results, and EKS98 nuclear shadowing are unable to reproduce the degree of suppression observed in the Au+Au data.  This implies that the collision conditions differ between p+p and heavy ion systems as is expected.  However, as shown in Fig.~\ref{raa_theory}a, final state suppression models without recombination included  generally over predict the suppression in central heavy ion collisions \cite{Capella} \cite{Grandchamp} \cite{Kostyuk}.  This is notable because these models have been found to agree well with the CERN SPS results \cite{NA50}.  Although, recent calculations by \cite{Zhu} reasonably match the PHENIX data.  Models which include quark recombination, Fig.~\ref{raa_theory}b, show a weaker degree of suppression and appear to reproduce the experimental nuclear modification \cite{Grandchamp} \cite{Bratkovskya} \cite{Andronic}.  The recombination models also predict a narrowing of the rapidity distribution of the $J/\psi$ invariant yield, $B\frac{dN}{dy}$, which is not observed in the data.  Figure~\ref{hi_jpsi_y} shows the $J/\psi$ invariant yield plotted as a function of rapidity at different collision centralities for $200~GeV$ Au+Au, Cu+Cu and p+p collisions.  No shape difference is evident between the data sets within errors.

\begin{figure}
\begin{center}
\includegraphics[width=4.0in]{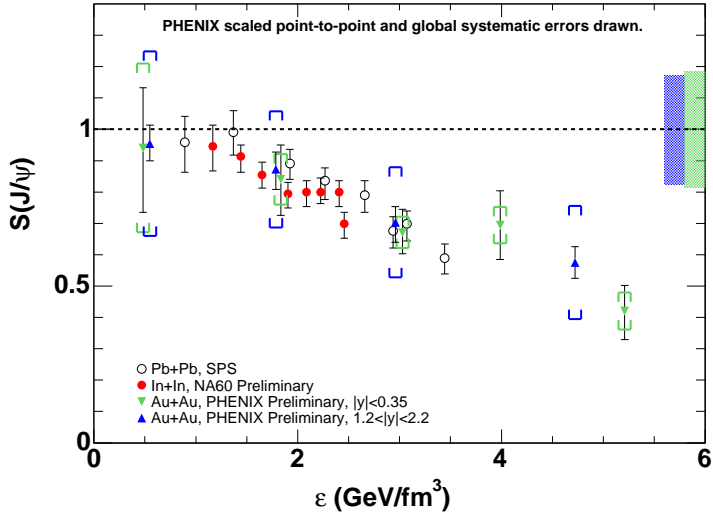}
\end{center}
\caption[satz_errors]{$J/\psi$ survival probability as a function of energy density.  The vertical error bars on the PHENIX data represent the statistical error, the brackets represent the point-to-point error and the global systematic is shown as a percentage error band around 1. \cite{satz_seq}}
\label{satz_errors}
\end{figure}

A comparison of the PHENIX heavy ion results with those obtained from Pb+Pb collisions by NA50 \cite{NA50} shows a similar degree of suppression in the nuclear modification factor as a function of energy density.  It has been proposed that this similarity is evidence of sequential melting of the charmonium states \cite{satz_seq}.  If the $J/\psi$ survival probability, $S(J/\psi)$, is taken as a measure of the relative production rate of $J/\psi 's$ compared to normal nuclear matter, then the data appear to indicate a maximum suppression of approximately $40\%$ in the most central collisions.  This is consistent with the feed-down contribution from the $\chi_{c}$ and $\psi'$ states which are expected to dissociate near the critical temperature while the $J/\psi$ can survive until approximately $2T_{C}$.  However, it is important to consider the current statistical and systematic errors on the PHENIX preliminary data shown in Fig.~\ref{satz_errors} which allow a large degree of latitude in this interpretation.  

\section{Other Quarkonia}
\begin{figure}
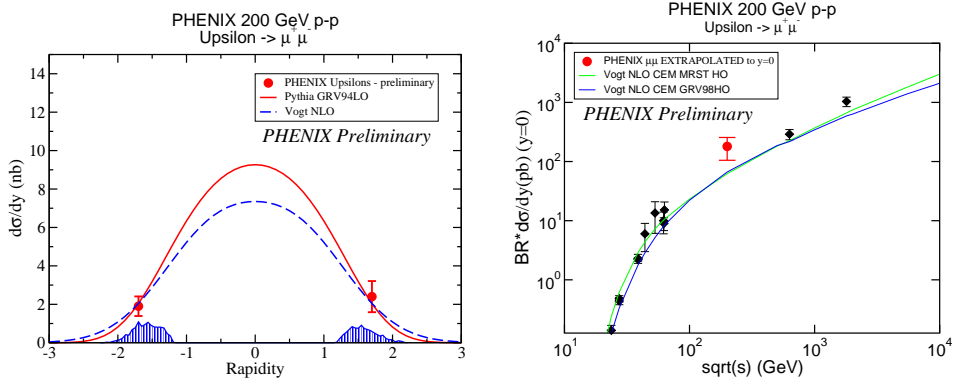

\begin{center}
$\begin{array}{c@{\hspace{0.25in}}c}
\includegraphics[width=2.4in]{ups_y.eps}&
\includegraphics[width=2.3in]{ups_roots.eps}
\end{array}$
\end{center}
\caption[upsilon]{$\Upsilon$ cross section in $200~GeV$ p+p collisions.} 
\label{upsilon}
\end{figure}

In addition to the $J/\psi$, the PHENIX detector has the capability of measuring the $\Upsilon$, $\chi_{c}$ and $\psi'$ quarkonia states in p+p collisions.   Shown in Fig.~\ref{upsilon} is the PHENIX preliminary $\Upsilon$ measurement as a function of rapidity and collision energy.  To constrain the shape of the rapidity dependence a measurement at mid-rapidity is required, but the preliminary cross section appears consistent with the trend observed in the worldÕs data.  The analysis of the $\chi_{c}$ and $\psi'$ states is still ongoing.  A factor of 3 improvement in the statistics available for these measurements is expected from the 2006 p+p run and the results will be forthcoming. 

\section{Summary}
While the PHENIX experiment has already produced an extensive set of quarkonia results, future detector upgrades and improved machine luminosity will allow these to be expanded upon.  Currently the p+p $J/\psi$ and $\Upsilon$ results are consistent with the world's data and provide a useful baseline reference for the comparison of the heavy ion results.  Improved measurements of $\Upsilon$, $\psi'$ and $\chi_{c}$ production are expected from existing p+p data sets.  Furthermore, the results obtained from d+Au collisions have begun to allow cold nuclear effects to be disentangled.  The PHENIX heavy ion results show striking similarities among collision species and energies.  Whether this is due to the recombination of uncorrelated quarks, or the sequential dissociation of charmonium states, or a yet to be determined mechanism remains to be resolved.  New and more precise future PHENIX measurements will shed further light on these processes and open additional avenues of understanding of quarkonia.



\end{document}